\documentclass{elsart}
\usepackage{graphicx,amssymb}
\journal{Physica A}
\begin{document}
\begin{frontmatter}
 
\title{Quantum Smoluchowski equation: Escape from a metastable state}
\author[iacs]{Dhruba Banerjee},
\author[bcb]{Bidhan Chandra Bag},
\author[iacs]{Suman Kumar Banik} and
\author[iacs]{Deb Shankar Ray\corauthref{cor}}
\corauth[cor]{Corresponding author, Tel.: +91 33 4733542, Fax: +91 33 4732805}
\ead{pcdsr@mahendra.iacs.res.in}
 
\address[iacs]{Indian Association for the Cultivation of Science,
Jadavpur, Calcutta 700 032, India}
\address[bcb]{Department of Chemistry, Visva-Bharati, 
Shantiniketan 731 235, India}
 
\begin{abstract}
We develop a quantum Smoluchowski equation in terms of a true
probability distribution function to describe quantum Brownian motion in
configuration space in large friction limit at arbitrary temperature and
derive the rate of barrier crossing and tunneling within an unified
scheme. The present treatment is independent of path integral formalism
and is based on canonical quantization procedure.
\end{abstract}
 
\end{frontmatter}

%%%%%%%%%%%%%%%%%%%%%%%%%%%%%% SECTION - I %%%%%%%%%%%%%%%%%%%%%%%%%%%%%%%%

\section{Introduction}

The theory of quantum Brownian motion is one of the major issues in
physics and chemistry today. Its tremendous success in the treatment
of various phenomena in quantum optics \cite{whl}, quantum tunneling and
coherence effects in condensed matter physics \cite{aoc,aocny,uw},
activated processes in chemical physics \cite{htb,pgw,whm,jl,jrc,bbbr}
is now well-documented in the current literature. While the noise operator
and density operator semigroup methods formed the core of development
in quantum optics in sixties and seventies, real time path integrals
attracted wide attention since eighties. However, these approaches are
plagued with several difficulties. First, a search for quantum analogue of
classical Fokker-Planck equation with a nonlinear potential, in general,
leads us to equations of higher (than two) derivatives of quasi-probability
distribution functions \cite{aoc,aocny,whz}. 
These distribution functions often become negative
or singular in the full quantum regime. Second, although large coupling
constants and large correlation times are treated nonperturbatively
formally in an exact manner by functional integrals their analytic
evaluation often poses severe difficulties as emphasized recently
by Stockburger and Grabert \cite{jts} and one has to consider special cases
and approximations, e.g., WKB or semiclassical limit. Moreover because
of well-known sign problem the numerical evaluation of the path integrals
are very difficult \cite{jts}. Third, in some situations \cite{apg}
the theory does not retain its validity as the
temperature $T \rightarrow 0$. This implies that vacuum fluctuations due to
heat bath are not correctly taken into account. The question is how to extend
classical theory to quantum domain for large friction at arbitrary
temperature in terms of a {\it true probabilistic description}. 
Very recently we have addressed \cite{bbbr} this issue to develop a 
non-Markovian quantum Kramers' equation describing Brownian motion in a
$c$-number phase space. The present analysis of overdamped limit is a brief
offshoot of this development in configuration space.
Our aim here is (i) to develop a
{\it quantum analogue} of classical Smoluchowski equation valid for
arbitrary temperature and (ii) to obtain the rate of escape from a
metastable state due to thermal activation and tunneling within an unified
description.

%%%%%%%%%%%%%%%%%%%%%%%%%%%%%% SECTION - II %%%%%%%%%%%%%%%%%%%%%%%%%%%%%%%

\section{The quantum Langevin equation}

To start with we consider the standard system-heat bath model whose 
Hamiltonian is given by
\begin{equation}
\label{eq1}
\hat{H} = \frac{\hat{P}^2}{2} + V(\hat{X}) +
\sum_j \left [ \frac{\hat{p}_j^2}{2} + 
\frac{1}{2} \kappa_j \left ( \hat{q}_j - \hat{X} \right )^2 \right ]
\end{equation}

\noindent
where $\hat{X}$ and $\hat{P}$ are the co-ordinate and momentum operators of 
the Brownian particle and the set
$\{ \hat{q}_j, \hat{p}_j \}$ is the set of co-ordinate and momentum
operators for the reservoir oscillators coupled linearly to the system
through the coupling constant $\kappa_j$ and obeying the usual commutation 
relation
$[\hat{X},\hat{P}] = i\hbar$ and $[\hat{q}_j,\hat{p}_j] = i\hbar \delta_{ij}$.
The potential $V(\hat{X})$ is due
to the external force field for the Brownian particle. 
Eliminating the bath degrees of freedom in the usual way \cite{whl}
we obtain the operator Langevin equation for the particle
\begin{equation}
\label{eq2}
\ddot{\hat{X}} (t) + \int_0^t dt' \; \beta (t-t') \; \dot{\hat{X}} (t')
+ V'(\hat{X}) = \hat{F} (t) \; \; ,
\end{equation}

\noindent
where the noise operator $\hat{F} (t)$ and the memory kernel $\beta (t)$
are given by 
\begin{equation}
\label{eq3}
\hat{F} (t) = \sum_j [ \{ \hat{q}_j (0) - \hat{X} (0) \} \kappa_j
\cos \omega_j t+\hat{p}_j (0) \kappa_j^{1/2} \sin \omega_j t] 
\; \; {\rm and}
\end{equation}

%\noindent
%and 
\begin{equation}
\label{eq4}
\beta (t-t') = \sum_j \kappa_j \cos \omega_j (t-t') \; \; ,
\end{equation}

\noindent
respectively, with $\kappa_j = \omega_j^2$.
Very recently \cite{bbbr} we have shown that on the basis of a quantum
mechanical average $\langle \ldots \rangle$ over the bath modes with
coherent states and the system with an arbitrary state Eq.(\ref{eq2}) can 
be cast into the form of a generalized quantum Langevin equation in
$c$-numbers,
\begin{equation}
\label{eq5}
\ddot{x} + \int_0^t dt' \beta (t-t') \dot{x} (t') + V' (x)
= f(t) + Q (x,t) \; \; .
\end{equation}

\noindent
In writing Eq.(\ref{eq5}) we denote the quantum mechanical mean value of
position operator $\langle \hat{X} \rangle$ = $x$ and
$Q (x,t) = V'(\langle \hat{X} \rangle ) - 
\langle V' ( \hat{X} ) \rangle $. $c$-number quantum Langevin force 
$f(t)$ satisfies $\langle f(t) \rangle_S = 0$ and
\begin{equation}
\label{eq6}
\langle f(t) f(t') \rangle_S = \frac{1}{2} \sum_j \kappa_j \hbar
\omega_j \left ( \coth \frac{\hbar \omega_j}{2 k_BT} \right )
\cos \omega_j (t-t') \; \; .
\end{equation}

\noindent
Here $\langle \ldots \rangle_S$ corresponds to an ensemble average
\cite{bbbr} over the quantum mechanical mean values of the 
co-ordinates and momenta of the bath oscillators.
Eq.(\ref{eq6}) is the celebrated quantum fluctuation-dissipation
relation \cite{aoc}. Quantum noise $f(t)$ and quantum fluctuation term
$Q (x,t)$ are due to heat bath and nonlinearity of the potential,
respectively. For details we refer to \cite{bbbr}.

%%%%%%%%%%%%%%%%%%%%%%%%%%%%%% SECTION - III %%%%%%%%%%%%%%%%%%%%%%%%%%%%%

\section{Large friction limit and quantum Smoluchowski equation}

We now consider the diffusion of a particle in an external potential
$V(x)$ as described by Eq.(\ref{eq5}). In the overdamped limit we drop
the inertial term $\ddot{x}$ and assume a Lorentzian density of modes of
heat bath oscillators such that
$ \kappa (\omega) \rho (\omega) = (2/ \pi)
[ \gamma / (1 + \omega^2 \tau_c^2) ]$, which in the limit
$\tau_c \rightarrow 0$ results in a damping kernel
$\beta (t-t') = \gamma \delta (t-t')$ in Eq.(\ref{eq5}). 
$\rho (\omega)$ refers to the density function used in the continuum limit.
Eq.(\ref{eq5}) then assumes the form
\begin{equation}
\label{eq7}
\dot{x} + \frac{1}{\gamma} V_{quant}' (x,t) = \frac{f(t)}{\gamma}
\end{equation}

\noindent
where we have expressed the effective quantum potential $V_{quant} (x,t)$
as $V_{quant}'(x,t)$ = $V'(x) - Q (x,t)$. Making use of the Stratonovich
prescription
the equivalent description of Eq.(\ref{eq7}) in terms of true probability
distribution $p (x,t)$ is given by
\begin{equation}
\label{eq8}
\frac{\partial p(x,t)}{\partial t} = \frac{1}{\gamma} 
\frac{\partial}{\partial x} \left [ V_{quant}' p(x,t) \right ] +
D_{qo} \frac{\partial^2 p}{\partial x^2} \; \; .
\end{equation}

\noindent
Here $D_{qo}$ is the quantum diffusion coefficient
which can be obtained with the following definition \cite{whl}
\begin{equation}
\label{eq9}
2 D_{qo} = \frac{1}{\Delta t} \int_t^{t+\Delta t} dt_1 
\int_t^{t+\Delta t} dt_2 \frac{1}{\gamma^2}
\langle f (t_1) f (t_2) \rangle_S
\end{equation}

\noindent
where the correlation function
$\langle f (t_1) f (t_2) \rangle_S$ of the $c$-number quantum noise of the
heat bath is given by Eq.(\ref{eq6}). Making use of Eq.(\ref{eq6})  in the 
continuum limit in Eq.(\ref{eq9}) we obtain
\begin{equation}
\label{eq9a}
2 D_{qo} = \frac{1}{2\gamma^2 \Delta t} \int_0^\infty d\omega
\kappa (\omega) \rho (\omega) \hbar \omega
\left ( \coth \frac{\hbar \omega}{2k_BT} \right ) I_\omega \; \; .
\end{equation}

\noindent
Here $I_\omega$ is given by
$I_\omega = \int_t^{t+\Delta t} dt_1 \int_t^{t+\Delta t} dt_2
\cos \omega (t_1 - t_2)$
which after explicit integration yields

\begin{equation}
I_\omega = \frac{4}{\omega^2} \sin^2 \frac{\omega \Delta t}{2} \; \; .
\end{equation}

\noindent
The general expression for quantum diffusion coefficient in the 
overdamped limit is therefore given by
\begin{equation}
\label{eq9d}
D_{qo} = \frac{2}{\gamma} \frac{1}{\pi \Delta t} \int_0^\infty
\frac{1}{1+\omega^2 \tau_c^2} \hbar \omega 
[ 2 \bar{n} (\omega) + 1 ]
\frac{1}{\omega^2} \sin^2 \frac{\omega \Delta t}{2} d\omega
\; , \; \tau_c \rightarrow 0
\end{equation}

\noindent
where we have put the Lorentzian density of states for bath oscillators
as stated earlier. $\bar{n} (\omega)$ is the average thermal photon number
of the heat bath and is given by
$\bar{n} (\omega) = 1/[\exp (\hbar \omega / k_BT) - 1]$.
The quantum diffusion coefficient can thus be 
calculated exactly by numerical integration over the bath frequencies
$\omega$ for $\tau_c \rightarrow 0$
(the Markovian limit). The following two limiting situations are 
further relevant for the present analytic treatment:
(i) In the high temperature limit $k_BT \gg \hbar \omega$,
the quantity $\hbar \omega [ 2 \bar{n} (\omega) + 1]$ in Eq.(\ref{eq9d})
reduces to $k_BT$ and the explicit integration results in 
$D_{qo} = k_BT /\gamma$, the usual Einstein's value in the static
friction limit. (ii) In the vacuum limit on the other hand
$\bar{n} (\omega) \rightarrow 0$ and $D_{qo}$ may be obtained by
considering the fact that the frequency dependence of the integrand in
(\ref{eq9d}) [or (\ref{eq9a})] except $I_\omega$ is flat (Markovian).
This results in $D_{qo} \simeq \hbar \tilde{\omega}/2\gamma$, where
$\tilde{\omega}$ is an average bath frequency which may be approximately
taken to be the linearized frequency of oscillation in the well
$\omega_0$ (i.e., $\tilde{\omega} \sim \omega_0$) since at $T \sim 0$
the dynamics is dominated by population at the bottom of the well. For
an arbitrary intermediate temperature we must rely, however, on the 
general expression (\ref{eq9d}) for the quantum diffusion coefficient.

\noindent
We now note that Eq.(\ref{eq8}) contains quantum corrections to all
orders. To make it more explicit we return to quantum mechanics of the
system in the Heisenberg picture to express the operators $\hat{X}$ and
$\hat{P}$ as
\begin{equation}
\label{eq11}
\hat{X} (t) = \langle \hat{X} (t) \rangle + \delta \hat{X} \; \; {\rm and}
\; \; \hat{P} (t) = \langle \hat{P} (t) \rangle + \delta \hat{P}
\end{equation}

\noindent
$\langle \hat{X} (t) \rangle$ and $\langle \hat{P} (t) \rangle$ are the
quantities signifying quantum averages and $\delta \hat{X}$ and
$\delta \hat{P}$ are quantum corrections. By construction
$\langle \delta \hat{X} \rangle$ and $\langle \delta \hat{P} \rangle$
are zero and they obey commutation relation
$[ \delta \hat{X}, \delta \hat{P} ] = i \hbar$. Using (\ref{eq11}) in
$V'(\hat{X})$ and a Taylor expansion around $\langle \hat{X} \rangle$ 
($\equiv x$) it is possible to express $Q(x,t)$ as
$Q(x,t) = - \sum_{n \geq 2} (1/n!) V_{n+1} (x)
\langle \delta \hat{X}^n (t) \rangle$ and $V_{quant} (x,t)$ as
\begin{equation}
\label{eq13}
V_{quant} (x,t) = V (x) +  
\sum_{n \geq 2} \frac{1}{n!} V_n (x) \langle \delta \hat{X}^n (t) \rangle
\end{equation}

\noindent
where $V_n(x)$ is the $n$-th derivative of the  
the classical potential which gets modified by the quantum
corrections. To solve quantum Smoluchowski equation (\ref{eq8})
it is therefore necessary to calculate the corrections
$\langle \delta \hat{X}^2 (t) \rangle$, 
$\langle \delta \hat{X}^3 (t) \rangle$, etc. To this end we return 
to operator Eq.(\ref{eq2}) and make use of the relation (\ref{eq11}) and a
Taylor expansion of the potential $V(\hat{X})$ to derive the following
coupled equations in the overdamped limit:

\begin{equation}
\label{eq14a}
\gamma \dot{x} + V'(x) + \sum_{n \geq 2} \frac{1}{n!} V_{n+1} (x) 
\langle \delta \hat{X}^n (t) \rangle = f(t)
\end{equation}

\begin{equation}
\label{eq14b}
\langle \delta \dot{\hat{X}^n} (t) \rangle = - \frac{n}{\gamma}
V''(x) \langle \delta \hat{X}^n (t) \rangle
\end{equation}

\noindent
where $n = 2, 3, \ldots$. A decisive advantage in the treatment of overdamped 
limit is noteworthy. The equations for quantum corrections
$\langle \delta \hat{X}^n (t) \rangle$ are closed in
contrast to those for the system without its surrounding \cite{akp,sm}. 
If one is interested in the local dynamics
around a point $x_0$ (say, at the bottom or top of a potential well)
the set of Eqs.(\ref{eq14b}) gets decoupled from (\ref{eq14a})
and one can obtain
simple estimates of $\langle \delta \hat{X}^n (t) \rangle$
since $V'' (x_0)$ can be treated as a constant in such cases.
More generally however,
Eq.(\ref{eq8}) can be combined with Eq.(\ref{eq14b}) to provide an
extended phase space description in terms of a true probability
function $p(x, \eta_2, \eta_3, \ldots, t)$
\begin{eqnarray}
\frac{\partial p(x, \eta_2, \eta_3, \ldots, t) }{\partial t}
& = & \frac{1}{\gamma} \sum_{n \geq 2} n \frac{\partial}{\partial \eta_n}
[ V''(x) \eta_n p ]
+ \frac{1}{\gamma} \frac{\partial}{\partial x} [ V'(x) p] 
\nonumber \\
& & + \frac{1}{\gamma} \frac{\partial}{\partial x} \left [
\sum_{n \geq 2} \frac{1}{n!} V_{n+1} (x) \eta_n p \right ]
+ D_{qo} \frac{\partial^2 p}{\partial x^2}
\label{eq15}
\end{eqnarray}

\noindent
where $\eta_n$ ($\equiv \langle \delta \hat{X}^n (t) \rangle$) for
$n = 2, 3, \ldots$ span the space of variables signifying quantum corrections
around quantum mechanical mean position $x$. 

We now discuss the classical and vacuum limits of quantum Smoluchowski
equation (\ref{eq8}). As mentioned earlier in the classical limit, 
$D_{qo}$ reduces to Einstein's classical diffusion coefficient
$k_BT/\gamma$. At the same time $Q(x,t)$ vanishes so that $V_{quant} (x,t)$
goes over to $V(x)$ and one recovers the usual classical Smoluchowski
equation. In the opposite limit as $T \rightarrow 0$, however, both quantum
noise due to system and vacuum fluctuation originating from the heat bath
make significant contribution. $D_{qo}$ in this limit assumes the form
$\hbar \omega_0/2\gamma$. In this context we refer to a recent
treatment \cite{apg} of large friction limit in quantum dissipative
dynamics to point out that the latter theory does not retain its full
validity as $T \rightarrow 0$ since the quantum noise due to heat bath
disappears in the vacuum limit.
Another noteworthy feature of quantum Smoluchowski equation (\ref{eq8})
[ or (\ref{eq15}) ] is that unlike Wigner function based equations \cite{whz}
it does not contain higher (than two) order derivatives of probability
distribution function. The positive definiteness of this function
is therefore ensured.

%%%%%%%%%%%%%%%%%%%%%%%%%%%%%% SECTION - IV %%%%%%%%%%%%%%%%%%%%%%%%%%%%%%

\section{Decay of a metastable state}

Based on the quantum Smoluchowski equation (\ref{eq8}) we now consider
the problem of escape from a metastable state. Over the last
two decades this has attracted a lot of attention in classical theory
of thermally activated processes \cite{htb}. Since a quantum analysis
naturally includes tunneling as an integral part of the problem we turn to 
this issue in search of an unified description of tunneling and 
thermal activation.

To begin with we first take into consideration of the time-scale
of free motion of the system, $\omega$ ($\omega^2 = V''$). 
Since according to Eq.(\ref{eq14b}) quantum correction
$\langle \delta \hat{X}^n (t) \rangle$ varies on a time-scale
$(n\omega^2/\gamma)^{-1}$ it is convenient to make an average of this
fluctuation over a period and replace $\langle \delta \hat{X}^n (t) \rangle$
by an average
$\overline{\langle \delta \hat{X}^n \rangle} = \omega
\int_0^{1/\omega} \langle \delta \hat{X}^n (t) \rangle dt$.
This implies quantum noise does not change significantly on a time-scale
during which the system relaxes so that $\tilde{V} (x)$ does not depend
on time explicitly. Eq. (\ref{eq8}) can then be recast
in the form of continuity equation
$ \partial p(x,t)/\partial t = - \partial J/\partial x$
where $J$ can be identified as a current which is given by
\begin{equation}
\label{eq17}
-J = \frac{1}{\gamma} \tilde{V}'(x) p + \frac{D}{\gamma}
\frac{\partial p}{\partial x} \; \; .
\end{equation}

\noindent
Here we have put
\begin{equation}
\label{eq18}
\tilde{V} (x) = V(x) + \sum_{n \geq 2} \frac{1}{n!} V_n (x)
\overline{\langle \delta \hat{X}^n \rangle}
\end{equation}

\noindent
and $D = \gamma D_{qo}$.
We now consider a cubic potential of the form $V(x)=
-(A/3) x^3 + B x^2$ ($A$ and $B$ are positive constants), 
with a metastable minimum at $x=0$ which is separated from the
true minimum by a finite potential barrier at $x = x_b$. Linearizing
the potential $V(x)$ at the barrier top we calculate the stationary 
flux $J$ ($\partial p/\partial t = 0$) around this point in the 
usual way,
\begin{equation}
\label{eq19}
J = \frac{\omega_b \sqrt{D}}{\sqrt{2\pi} \gamma} p(0)
\exp (-E/D) \exp \left [ \frac{-(1/2) V''' (x_b) 
\overline{ \langle \delta \hat{X}^2 \rangle }_b x_b}{D} \right ]
\end{equation}

\noindent
where $\omega_b$ refers to the frequency at the barrier top and $E$ is
the activation energy ($E=V(x_b)$).
Again considering an equilibrium distribution or zero current situation
($J=0$) around $x=0$ we linearize the potential
$V(x)$ and derive from Eq.(\ref{eq17}), the population in the well as
given by
\begin{equation}
\label{eq20}
n_a = \frac{\sqrt{2\pi D}}{\omega_0} p(0)
\end{equation}

\noindent
where $\omega_0$ refers to the frequency in the well at $x=0$.
Based on the reasoning given earlier, 
we can obtain an estimate of quantum correction in (\ref{eq19}) as
$\overline{\langle \delta \hat{X}^2 \rangle}_{x=x_b} =
(\hbar / 2 \omega_b) [ 1 + (2 \omega_b / \gamma) ]$ which
follows from the solution of Eq.(\ref{eq14b}) and a quantum average with
minimum uncertainty state
$\langle \delta \hat{X}^2 (0) \rangle_{x=x_b} = (\hbar / 2 \omega_b)$.
Furthermore $x_b$ may be expressed as $x_b = \sqrt{2E}/\omega_0$. 
With these simplifications the rate, $k$ is given by the ratio
$J/n_a$ :
\begin{equation}
\label{eq22}
k  =  \frac{\omega_0 \omega_b}{2 \pi \gamma} \exp (-E/D)
        \exp \left [
\frac{  (\sqrt{2E}/\omega_0) (\hbar A/2 \omega_b)
\left ( 1 + \frac{2\omega_b}{\gamma} \right ) }{D}
\right ] \; \; .      
\end{equation}

The expression (\ref{eq22}) is the
quantum rate of decay of a metastable state at any temperature. 
The quantum feature appears in two different ways. First,
since $D$ may
be treated as a quantum analogue of $k_BT$ in the Boltzmann exponential
factor, which reduces to $\hbar \omega_0/2$ in the vacuum limit
and to $k_BT$ in the thermal limit, the quantum nature of the heat bath
is manifested through $D$. Second, the quantum nature of the system is
pronounced through quantum correction terms which are entangled with the
nonlinearity of the potential and vanishes in the classical limit. Thus in
the classical limit (\ref{eq22}) reduces to the familiar expression
for the Kramers'-Smoluchowski rate of escape \cite{hak},
\begin{equation}
\label{eq25}
k_{cl}  =  \frac{\omega_0 \omega_b}{2 \pi \gamma} \exp (-E/k_BT) \; \; .
\end{equation}

\noindent
In the vacuum limit, i.e., at $T \sim 0$, on the other hand we have
\begin{equation}
\label{eq26}
k_{vac}  =  \frac{\omega_0 \omega_b}{2 \pi \gamma}
\exp (-2E/\hbar \omega_0 ) \exp \left [
\frac{ A (\sqrt{2E}/\omega_0) (\hbar / 2\omega_b)
\left ( 1 + \frac{2\omega_b}{\gamma} \right )
}{
(1/2) \hbar \omega_0
}
\right ] \; \; .
\end{equation}

\noindent
The vacuum contribution corresponds to quantum tunneling at zero
temperature. 
It is important to recall that the problem of zero temperature
tunneling was first successfully addressed by Caldeira and Leggett 
\cite{aoc,aocny} in early eighties which subsequently initiated a major
advancement in the field of macroscopic quantum tunneling \cite{htb}.
The main results in the context of overdamped limit can be summarized
as follows: First, the damping causes an exponentially strong
suppression of tunneling rate. Second, for strong dissipation there is
a large region where thermal and quantum fluctuations interplay so that
the total rate of decay increases because the thermal activation is 
supplemented by quantum tunneling at finite temperature. These features
can be easily recovered in the present theory. The rate coefficient in
the vacuum limit $k_{vac}$ shows that as the temperature approaches zero
tunneling decay is almost exponential. Furthermore, at finite temperature,
the presence of factor $(\hbar A)/(D \omega_b)$ in the exponential in the
total rate coefficient in Eq.(\ref{eq22}) implies an interplay of thermal
fluctuation in $D$ and quantum fluctuation due to $(\hbar A)/\omega_b$
arising out of nonlinearity of the system. This leads to a effective 
reduction of the activation barrier from its classical value $E$ resulting
in an increase of the rate of decay. It is thus apparent that although the
results are qualitatively comparable the underlying physical picture
in the two schemes (the path integral and the present one) are
different. The tunneling decay at low temperature in the formal case is
primarily determined by the nature of bounce solution evaluated by a 
semiclassical
steepest decent method which depends on the effective path between the
turning points. At the high temperature (i.e., in the classical regime)
an unified multidimensional WKB approach is generally
advocated to recover classical result \cite{uw,htb}. The classical theory
in the present formulation on the other hand is contained in $k_{cl}$
(Eq.(\ref{eq25})) which shows that the rate of thermal activation
varies inversely with damping constant according to 
Kramers'-Smoluchowski limit.
The quantum rate coefficient (\ref{eq22}) thus interpolates between
high temperature thermal activation and zero temperature tunneling.

%%%%%%%%%%%%%%%%%%%%%%%%%%%%%% CONCLUSION %%%%%%%%%%%%%%%%%%%%%%%%%%%%%%%%

\section{Conclusions}

In conclusion, we have considered the strong friction limit in quantum
stochastic processes to develop a quantum Smoluchowski equation.
The rate of decay of a metastable state which includes thermal 
activation and tunneling within a single scheme has been derived. Since
classical Smoluchowski equation has a diverse range of applicability,
e.g., in fluctuating barrier \cite{dg}, in thermal ratchet
and molecular motors \cite{mom,rda,jap}, in tunnel diodes \cite{rnm}
and in the decay of periodically driven metastable potentials 
\cite{sdg,lrh,dls} we hope the present quantum analogue
to keep its potential in similar issues.

This work was supported by the Council of Scientific and Industrial Research
(C.S.I.R.), Government of India, under grant No. 01/(1740)/02/EMR-II.

%%%%%%%%%%%%%%%%%%%%%%%%%%%%%% REFERENCES %%%%%%%%%%%%%%%%%%%%%%%%%%%%%%%%

\end{document}